\documentclass[12pt]{article}

\usepackage{amsmath, amsthm, amssymb}

\usepackage[russian,english]{babel}
\let\oldmarginpar\marginpar
\renewcommand\marginpar[1]{\-\oldmarginpar[\raggedleft\small\sf #1]%
{\raggedright\small\sf #1}}

\usepackage{ulem} % to define \sout{ }
\title{\bf The Einstein formula: $\mbox{\boldmath $E_0=mc^2$}$.
\\``Isn't the Lord laughing?''\thanks{Published: Uspekhi
Fiz. Nauk {\bf 178} (5) (2008) 541-555 (in Russian);
Physics--Uspekhi {\bf 51} (5) (2008) 513-527 (in English).}}

\author{L.B. Okun  \\ ITEP, Moscow, Russia}

\date{}

\begin{document}

\maketitle

\abstract The article traces the way Einstein formulated the
relation between energy and mass in his work from 1905 to 1955.
Einstein emphasized quite often that the mass $m$ of a body is
equivalent to its rest energy $E_0$. At the same time he
frequently resorted to the less clear-cut statement of equivalence
of energy and mass. As a result, Einstein's formula $E_0=mc^2$
still remains much less known than its popular form, $E=mc^2$, in
which $E$ is the total energy equal to the sum of the rest energy
and the kinetic energy of a freely moving body. One of the
consequences of this is the widespread fallacy that the mass of a
body increases when its velocity increases and even that this is
an experimental fact. As wrote the playwright A N Ostrovsky
``Something must exist for people, something so austere, so lofty,
so sacrosanct that it would make profaning it unthinkable.''
%\tableofcontents

\bigskip

{\bf Contents}

~~~~~~

1 \; Introduction \hfill 2

2 \; Prologue. The years 1881--1904 \hfill 2

3 \; 1905 --- annus mirabilis \hfill 3

4 \; Have I been led around by the nose? \hfill 4

5 \; 1906 -- 1910. Minkowski \hfill 5

6 \; 1911 -- 1915. On the road to General Relativity Theory \hfill
6

7 \; 1917. Cosmological constant \hfill 8

8 \; 1918 -- 1920. Noether \hfill 8

9 \; 1921. ``The Meaning of Relativity'' \hfill 9

10 \; 1927 -- 1935 \hfill 10

11 \; 1938 -- 1948. Atomic bomb \hfill 12

12 \; 1949. Feynman diagrams \hfill 16

13 \; 1952 -- 1955. Last years \hfill 17

14 \; Born, Landau, Feynman \hfill 17

15 \; Epilogue \hfill 19

16 \; Conclusion \hfill 19

Postscriptum. In memory of J A Wheeler \hfill 20

Acknowledgements \hfill 20

References \hfill 20

\section{Introduction}

The formula $E=mc^2$ is perhaps the most famous formula in the world. In the
minds of
hundreds of millions  of people it is firmly associated with the menace of atomic
weapons. Millions perceive it as a symbol of relativity theory. Numerous authors
popularizing science keep persuading their readers, listeners, and viewers that
the mass of any body (any particle) increases, as prescribed by this formula, when
its velocity
increases. And only a small minority of physicists --- those who
specialize in elementary particle physics --- know that Einstein's true
formula is $E_0=mc^2$, where $E_0$ is the energy contained in a body at rest,
and that the mass of a body is independent of the velocity at which it travels.

Most physicists familiar with special relativity know that in it,
the energy $E$ and momentum ${\bf p}$  of a freely moving body are
related by the equation $E^2-{\bf p}^{2}c^2=m^2c^4$ where $m$ is
the mass of the body. Alas, not all of them realize that this
formula is incompatible with $E=mc^2$. But an even smaller number
of people know that it is perfectly compatible with $E_0=mc^2$,
because $E_0$ is the value assumed by $E$ when ${\bf p} = 0$. This
article is written for those who do not want to be lost in three
pines\footnote{``To be lost in three pines'' in Russian is
equivalent to ``loose one's way in broad daylight'' in English.
({\it Author's note to English version of the article.})} of the
above three formulas and who wish to attain a better understanding
of relativity theory and its history.

When Einstein first introduced the concept of rest energy in 1905 and discovered
that the mass of a body is a measure of the energy contained in it, he felt so
amazed that he wrote in a letter to a friend: ``for all I know, God Almighty
might be laughing at the whole matter and might have been leading me around by
the nose.'' In what follows, we see how throughout his life Einstein
returned again and again to this same question.

We shall see how the formula $E_0=mc^2$ made its way through Einstein's
writings. Also, how he carefully emphasized that the mass of a body depends on
the amount of energy it contains but never stated (in contrast to his
popularizers!) that mass is a function of the body's velocity. Nevertheless, it
is true that he never once rejected the formula $E=mc^2$ that is believed to be
`his formula' and in the mass psyche is an icon of modern physics.

To my reader: If you feel bored with following the meticulous analysis and collation of
texts, please jump to the Epilogue and the adjacent sections, where I
have tried to briefly describe the results of the
analysis without going into technicalities. It is possible that after you do so,
reading about Einstein's many attempts to clarify the relation between energy and mass will become
more interesting and compelling.

When writing this review, I used Einstein's historically first ever collected
works [1]. (This four-volume edition was published in Russian in
1965 -- 1967.) Where possible, I also used the multivolume Princeton Collected
Papers. (Volumes with `all papers and documents' by Einstein [2] and
their translations into English [3] began to appear in Princeton in
1987. In 2007, ten volumes were published, of which five (1, 5, 8, 9, 10) contain his
correspondence until 1920 and five (2, 3, 4, 6, 7) contain his works until 1921).

\section{Prologue. The years 1881--1904}

It is well known that the principle of relativity dates back to Galileo [4] and
Newton [5], and that the theory of relativity was constructed in the papers of
Lorentz, Einstein, Poincar\'e, and Minkowski [6].

The notion of velocity-dependent mass was born in the years preceding the creation of the
theory of relativity and in the first years after its creation.

It was molded in the papers of Thomson [7], Heaviside [8],
Searle [9], Abraham [10], and also Lorentz
[11] and Poincar\'{e} [12], who tried hard to
have Maxwell's equations of electromagnetism to
agree
with the equations
of Newton's mechanics.
These publications stimulated the experiments of
Kaufmann [13] and Bucherer [14, 15].
They used formulas
of Newton's nonrelativistic mechanics to process their experimental data and
concluded that mass increases with increasing velocity.

It was the matter not only of formulas as such but also of the
very spirit, the very foundations of the nonrelativistic physics
in which mass is a measure of inertia of a body. It was difficult
to comprehend, at the borderline between the 19th and 20th
centuries, that these foundations were being replaced by a more
general base: the measure of inertia of a body is not its mass but
its total energy $E$ equal to the sum of rest energy and kinetic
energy. The fact that
%kinetic
energy $E$
%had
entered with a factor $1/c^2$ prompted
people
to interpret
$E/c^2$ as the mass. In fact the progress in relativity theory,
%mostly
achieved mostly %by
through the efforts of Einstein, Minkowski, and Noether, showed
that it was necessary to connect mass not with total energy but
only with rest energy.

\section{1905 --- annus mirabilis} %vik

In 1905, Einstein published his three ground-breaking, fundamental papers dealing with
the properties of light and matter [16 -- 18].

In [16], he introduced the concept of the quantum of energy of
light and, using this concept, explained the photoelectric effect,
which had been experimentally discoverednot long before that. (The
value of the Planck constant $h$ --- the quantum of action --- had
been established earlier, see [19].)

In [17], Einstein considered almost the entire set of consequences of the
principle of relativity and of the finite speed of light. Thus he
derived in \S 8 the
formula for the transformation of the energy of light in the transition from one inertial
reference frame to a different one that moves at a velocity $v$ relative to the former:

$$ \frac{E'}{E}\ = \frac{ 1-(v/V)\cos\phi}{\sqrt{1-(v/V)^2}}.$$

Here $V$ is the velocity of light and $\phi$ is the angle between the direction of motion
of light and that of the observer. Then in \S 10 he obtained the expression for the
kinetic energy of the electron:

$$ W=\mu V^2\left(\frac{1}{\sqrt{1-(v/V)^2}}-1\right) \;, $$

where $\mu$ is the mass of the electron and $v$ is its velocity.

(Furthermore, in \S 10, Einstein derived expressions for the so-called longitudinal $m_l$ and transverse %al
$m_t$ masses of the electron that Abraham and Lorentz had earlier introduced
%\cite{lore1904}
and
he obtained:
\begin{eqnarray}
m_l &=& \frac{\mu}{(\sqrt{1-(v/V)^2})^3}\, ,\nonumber \\
m_t &=& \frac{\mu}{1-(v/V)^2}.\nonumber
\end{eqnarray}

The second of these expressions %that
differs from Lorentz's $m_t$
and
is wrong, and later
Einstein never insisted on it.)

As regards the formulas for the kinetic energy $W$ of an electron and for a photon
energy $E'$, he applied both these formulas in the next paper [18]
when deriving the relation between mass and energy.

There he
considered `two amounts of light,' with energy $L/2$ each, both emitted by a
massive body at rest but traveling in opposite directions. In this paper,
Einstein for the first time introduced the rest energy of a massive body,
denoting it by $E_0$ before emission and by $E_1$ after.
In view of the energy conservation law,

$$ E_0 - E_1 = L .$$

He then looked at the same process in a reference frame moving at a velocity $v$ relative
to the body, and obtained the following expression for the difference between kinetic
energies of the body before and after the act of emission:

$$ K_0 - K_1 = L\left(\frac{1}{\sqrt{1-(v/V)^2}} -1\right) \; .$$

He also specially pointed out that the difference between kinetic energies contains an
arbitrary additive constant $C$ included in the expression for energy. He returned to the
matter of the constant $C$ many times during the subsequent 50 years; we discuss it
later in this paper.

The left- and right-hand sides of the equality above depend on  $v$ in the same manner,
as follows from the expression for $W$. Since the velocity  $v$ is the same before and
after the emission, while the kinetic energy of the body decreased, this immediately
implies that the mass of the body decreased by the amount $L/V^2$. From this Einstein
concluded that ``The mass of a body is a measure of its energy content'' and remarked that
it might be possible to check this conclusion in the decays of radium.

The title of the paper is noteworthy: ``Does the inertia of a body
depend on its energy content?''. Considered together with the
contents of the paper, it indicates that it was the mass that
Einstein identified with the measure of a body's inertia. But this
is only valid in Newton's approximation. As we know today, the
measure of a body's inertia in relativity theory is its total
energy E: the greater the total energy of a body, the greater its
inertia. (By the `measure of a body's inertia,' we here mean the
proportionality coefficient between momentum and velocity. There
is no universal proportionality coefficient between force and
acceleration in relativity theory. Lorentz and Abraham had already
established this when they introduced the longitudinal and
transverse masses.)

Einstein held to the opinion that the energy of a free body is defined in
relativity theory only up to an additive constant, by analogy to potential
energy in Newtonian mechanics. This may have resulted in his underestimating his own
revolutionary step forward --- the introduction of the concept of rest
energy into physics.
There is nothing special about the rest energy $E_0$ once energy is only defined up to $C$.

But as we know today, there is no place for $C$ in the theory he created.
The energy and momentum of a free particle are uniquely defined in the theory by
the relation $E^2-{\bf p}^2c^2=m^2c^4$; we return to it more than once in what follows.

\section{Have I been led around by the nose?}

The discovery that mass depends on energy struck Einstein so forcibly that he
wrote in a letter to his friend Conrad Habicht [20] (see also [21]):

``A consequence of the study on electrodynamics did cross my mind.
Namely, the relativity principle, in association with Maxwell's fundamental
equations, requires that the mass be a direct measure of the energy contained in a
body; light carries mass with it. A noticeable reduction of mass would have to take
place in the case of radium. The consideration is amusing and seductive; but for
all I know, God Almighty might be laughing at the whole matter and might have been
leading me around by the nose.''

It looks as if God continues to lead the interpreters of the relativity theory by the
nose much as He did in Einstein's time.

\section{1906 -- 1910. Minkowski}

\subsection*{1906}

In 1906, Einstein published two papers on relativity theory: [22, 23].
In [22], he treated mass transfer by light in a
hollow cylinder from its rear face to the front. For the cylinder not to move
as a whole, he imposed the condition that light with an energy $E$ has the mass
$E/V^2$; he thereby reproduced Poincar\'{e}'s result of 1900 [12].
Presumably, he considered it inadmissible for the energy and mass carrier to have zero
mass (to be massless). In [23], he considered a method for
determining the ratio of longitudinal and transverse masses of the electron
previously introduced by Lorentz and Abraham.
As far as mass is concerned,
therefore, these papers were a step back in comparison with [18].

\subsection*{1907}

In 1907, Einstein published four papers on relativity theory: [24
-- 27]. The first of these discussed the frequency of radiation
from an atom. The second emphasized the difference between the
relativity principle and the relativity theory. He considered his
own work as dealing with the principle of relativity, which he
regarded as being analogous to those of thermodynamics. As for the
theory of relativity, he believed that it was yet to be
constructed.

The paper that is especially significant for us here is [26],
which gave the formulation of the mass--energy equivalence
(see footnote in \S 4):
``One should note that the simplifying assumption  $\mu V^2= \varepsilon_0$ is also
the expression of the principle of the equivalence of mass and energy ...''
(The simplifying assumption referred to here is the choice of an arbitrary constant in the
expression for energy.)

The most detailed among the papers published in 1907 was [27]. It
consists of five parts: (1) Kinematics (\S 1 -- \S 6). (2)
Electrodynamics (\S 7). (3) Mechanics of a material point
(electron) (\S 8 -- \S 10). (4) On the mechanics and
thermodynamics of systems (\S 11 -- \S 16). (5) Relativity
principle and gravitation (\S 17 -- \S 20). Short note [28] with
corrections of misprints and elaborations belongs to this group of
papers.

Of special interest for us are parts 4 and 5. In part 4, Einstein discussed the
additive constant in the energy and showed that it is not included in the
relation between momentum, energy, and velocity of a body. Part 5 ended with the
following words:
\begin{quote}
``Thus the proposition derived in \S 11, that to an amount of energy $E$ there corresponds
a mass of magnitude $E/c^2$, holds not only for the {\em inertial} but also for the
{\em gravitational} mass, if the assumption introduced in \S 17 is correct.''
\end{quote}

On
the
one hand, this sentence states that energy, not mass, is both the
measure of inertia and the source of gravitation. But on the other hand, it
can be understood to say that a photon with an energy $E$ has both the inertial
mass
and the gravitational
mass equal to $E/c^2$. This ambiguous interpretation
continues to trigger heated debates.

\subsection*{1908} % Minkowski } %vik

In 1908, Einstein together with J Laub published two articles on the
electrodynamics of moving macroscopic bodies: [29, 30] (see
also [31, 32].) Although pertaining to
relativity theory, these papers are nevertheless not relevant to the problem under
discussion here, the relation between energy and mass.

The talk delivered by Hermann Minkowski in 1908 [33] was an important
milestone in the history of relativity theory. Minkowski was the first to propose the
four-dimensional spacetime formulation of the theory. In this formulation, as we know, the
mass of a particle is a quantity independent of its velocity.

It may seem paradoxical but the first paper by Lewis [34] declaring that
the mass equals $E/c^2$ appeared at the same time. This standpoint was further developed
and spread by Lewis and Tolman in [35 -- 38].

\subsection*{1909}

Einstein's paper [39] published in 1909 is not concerned with the
relation between mass and energy.
But we find a number of statements in his articles
[40 -- 42] published at the same time
that shed much light on his understanding of this problem. For instance, in
[42], which contains the text of Einstein's first public speech (at a
congress of German natural scientists in Salzburg), he wrote:

``The first volume of the excellent textbook\footnote{``The Course of Physics'' by
O D Chwolson (volumes 1 and 2) was published in Russian in 1897;
its German translation appeared in 1902.} by Chwolson which was published in
1902, contains in the Introduction the following sentence about the ether: `The
probability of the hypothesis on the existence of this agent borders
extraordinarily closely on certainty.' However, today we must regard the ether
hypothesis as an obsolete standpoint.''

Then:
``\ldots the inertial mass of a
body decreases upon emission of light\ldots Energy and mass appear as equivalent
quantities the same way that heat and mechanical energy do\ldots The theory of
relativity has thus changed our views on the nature of light insofar as it does
not conceive of light as a sequence of states of a hypothetical medium but
rather as something having an independent existence just like matter.''

\subsection*{1910}

In 1910, A Einstein and L Hopf discussed the application of probability theory
to
the analysis of
the properties of radiation [43, 44].

At the same time, Einstein published in a French journal a
major review of relativity
theory [45] devoted mostly to the
transformations of
spatial coordinates
and time but also briefly outlining Minkowski's ideas about the four-dimensional
world. Only at the
end of this paper
did he mention that
\begin{quote}
``\dots the mass of any arbitrary body
{\em depends on the quantity of energy} it contains\dots Unfortunately, the change of
mass $W/c^2$ is so slight that one cannot hope for its detection by experiment for the time being.''
\end{quote}
Einstein did not stipulate that by ``energy $W$ contained in a
body'' he meant rest energy.

\section{1911 -- 1915. On the road to General Relativity Theory} %6} %vik

\subsection* {1911}

In 1911, Einstein published three papers on the theory of relativity:
[46 -- 48].

In [46], he discussed the propagation of light in a gravitational
field, starting with the assumption that a photon with energy $E$
has an inertial and a gravitational mass, both of which are equal
to $E/c^2$, and he calculated that the angle of deflection of
light by the Sun's gravitational field would be 0.83 arc second
--- which is half the correct value that he would later derive (in
1915) using general relativity.

(I should remark that the same ``half value'' had already been obtained and published by
Soldner in 1804 (see [49, 50]). But Einstein was not aware of it:
Soldner's paper was totally forgotten soon after its publication.)

At the end of review paper [47] devoted mostly to clocks and rods
in relativity theory, Einstein mentioned uniting the law of
conservation of mass with the law of conservation of energy:
``However odd this result might seem, still, in a few special
cases, one can unequivocally conclude from empirically known
facts, and even without the theory of relativity, that the
inertial mass increases with energy content.'' Perhaps this
sentence refers to experiments of Kaufmann and Bucherer. But this
would suggest that he believed that mass increases with increasing
kinetic energy and therefore with increasing velocity.

A short note [49] discussed the contraction of the length of a moving rod.

\subsection*{1912}

Einstein's papers of this period
[51 -- 55] were mostly attempts to
create a more general relativity theory that would embrace gravitation. Only lectures
[51] dealt with special relativity.

His statements made during 1912 again display the above-mentioned ambiguity in the
interpretation of mass as the equivalent of rest energy, on the one hand, and as a measure of
inertia, on the other.

We find there a statement that $m$ should be considered %as
to be a characteristic constant of a
`material point' (massive point-like body), which
does not vary as a function of the object's motion. On the other hand, it is also
stated that the energy of a free particle is defined only up to
an arbitrary additive
constant. Nevertheless, $mc^2$ equals the rest energy (see the discussion of equation
%(28')
$(28')$.)

\subsection*{1913--1914}

In paper [56] co-authored with M Grossmann, Einstein continued to
discuss the proportionality between the inertial and gravitational
masses, which had been measured with high accuracy in experiments
by E\"{o}tv\"{o}s, and he discussed the dependence of the speed of
light c on the gravitational potential.

In 1914, Einstein published a short note expounding his point of view
regarding
the concept of mass [57]. A manuscript with a synopsis of his lectures on
special relativity theory dates back to the same period [58].

In [57], he discussed the contribution of the gravitational field
to the gravitational and inertial masses of a body and came to
the conclusion
that the inertia of a closed system is entirely determined by its rest energy.

Paper [58] gave an expression for the energy -- momentum 4-vector
and the relation $E_0/c^2=m$ which would appear again only in
1921. We note that $m$ was referred to in [58] as rest mass
(Ruhemasse), which seems to imply that the mass of a body at rest
is not the same as when the body moves.

\subsection*{1915}

The year 1915 was marked by the completion of general relativity theory, in paper
[59]. In fact, already in his preceding paper [60], Einstein had derived
formulas that described two
most important
effects of this theory: the precession of Mercury's
perihelion and the deflection of light by the gravitational field of the Sun. The secular
motion of Mercury's perihelion (about $40''$ per century), which could not be explained in
terms of the influence of the known bodies in the solar system, was established by
Le Verrier in 1859. Einstein calculated that general relativity theory predicted secular
precession as $43''$.

But the true world fame came from prediction of the angle of deflection of
light by $1.7''$ after it had been confirmed by the British expedition that
observed the solar eclipse in 1919.

\section{1917. Cosmological constant}

A book was published in 1917 to popularize relativity theory [61].
It dealt mostly with the joint transformation of space and time
coordinates. However, $\S 15$ mentioned the kinetic energy of a
material point, which now equaled not $mv^2/2$ but $mc^2/\sqrt {1
-v^2/c^2}$, and therefore incorporated both its kinetic energy
proper and its rest energy. Then we read this:

``Before the advent of relativity, physics recognized two conservation laws of
fundamental importance, namely, the law of the conservation of energy and the law
of the conservation of mass; these two fundamental laws appeared to be quite
independent of each other. By means of the theory of relativity they have been
united into one law.''

And even though an attentive reader
concludes from the text that follows that
Einstein was speaking %spoke
of $E_0=mc^2$, a slightly less attentive reader %may
might %decide
guess that
$E=mc^2$ was meant.
The fact that at times Einstein treated rest energy as part of kinetic energy did not help
to clarify matters.

The most famous among Einstein's papers published in 1917 was called
``Cosmological Considerations on the General Theory of Relativity''
[62]. There Einstein formulated for the first time the possibility of a
non-vanishing energy density of the vacuum; he denoted it by the letter
$\lambda$. This energy density is the same at every point in the Universe. It is
essentially a completely delocalized energy, spread over the entire Universe.

Einstein introduced this cosmological constant --- the
so-called $\lambda$-term --- in order to be able to describe a stationary Universe in
general relativity. It soon became clear, however, that a stationary solution cannot be achieved
in this manner.

In 1922, Friedmann, while reading this paper by Einstein, advanced his theory
of the expanding Universe [63, 64]. Einstein first dismissed Friedmann's
arguments [65], but then accepted them [66]. In 1929, Hubble published the
first observational data [67] supporting the expansion of the Universe.

In 1945, Einstein published the second edition of his book ``The Meaning of Relativity''
with a special addendum ``On the Cosmological problem'' devoted to the theory of
the expanding Universe [68].
At the turn of the 1970s -- 1980s, a model of the exponentially fast
expansion (inflation) of the early Universe was suggested [69 -- 71].
According to this model, the effective cosmological term forms when the Universe is created,
due to a nonzero
mean vacuum value of a special scalar field, which later transforms into
high-energy particles.

In 1998 -- 1999, two groups of observers measuring the luminosity and spectra of
supernovas came to a conclusion that the rate of the expansion of the Universe
is increasing [72, 73] (see also [74].)
The available data indicate that ordinary
matter contains only 4\% of the energy of the Universe, that about 24\%
is contained in the particles of the so-called dark matter whose nature is as yet unknown, and
about 70\%
of the entire energy of the Universe
is usually referred to as dark energy and attributed to
Einstein's cosmological constant $\lambda$ .

\section{1918 -- 1920. Noether}

\subsection*{1918}

In 1918, the brilliant paper of Emmy Noether was published [75], in which
she proved, among other things,
that the dynamic conservation laws are implied by the symmetry properties of space -- time. We
know that conservation of energy is a consequence of the uniformity of time, and that
conservation of momentum is a consequence of the uniformity of space. Angular momentum is conserved as a
result of the isotropy of space: physics remains unchanged if coordinate axes undergo rotation
in the planes $xy$, $yz$, $zx$.
Similarly, Lorentz invariance follows from the fact that physics remains unchanged
under pseudo-Euclidean rotations in the planes $xt$, $yt$, $zt$.
Einstein wrote very enthusiastically about this discovery of
Noether in a letter to Hilbert [76]:

``Yesterday I received a very interesting paper by Ms. Noether about the
generation of invariants. It impresses me that these things can be surveyed
from such general point of view. It would not have harmed the G\"{o}ttingen old
guard to have been sent to Miss Noether for schooling. She seems to know her trade well!''

Soon after that Einstein sent for publication a paper [77] on the
conservation of energy in general relativity, which presented a statement that the
energy of a closed system plays the role of both inertial and gravitational
mass.

\subsection*{1919}

Among the publications of 1919, I need to specially mention a
short note ``A test of the general theory of relativity'' [78] on
the discovery of the deflection of light rays by attraction of the
Sun and an article in {\it The Times}  entitled ``What is the
theory of relativity?'' [79]. Among other things, Einstein wrote:

``The most important upshot of the special theory of relativity
concerned the inertial masses of corporeal systems. It turned out
that the inertia of a system necessarily depends on its
energy-content, and this led straight to the notion that inert
mass is simply latent energy. The principle of the conservation of
mass lost its independence and became fused with that of the
conservation of energy.''

\subsection*{1920}

In 1920, Einstein prepared a draft manuscript of an extensive
popular article ``Fundamental ideas and methods of the theory of
relativity, presented in their developments.'' Einstein worked on
this article as an invited publication in {\it Nature}, but it was
never published [80].

At the same time, Einstein's letter appeared in a Berlin newspaper, ``My response.
On the anti-relativity company'' [81].
The letter opens with the words:

``Under the pretentious name ``Arbeitsgemeinschaft deutscher Naturforscher,'' a
variegated society has assembled whose provisional purpose of existence
seems to be to degrade, in the eyes of nonscientists, the theory of relativity
as well as me as its originator.''

Then Einstein wrote:
``\ldots I have good reasons to believe that motives other than the striving
for truth are at the bottom of this business. [\ldots] I only answer because
well-meaning circles have repeatedly urged me to make my opinion known.

First, I want to note that today, to my knowledge, there is hardly a scientist
among those who have made substantial contributions to theoretical physics
who would not admit that the theory of relativity in its entirety is founded
on a logical basis and is in agreement with experimental facts which to date
have been reliably established. The most important theoretical physicists --- namely,
H A Lorentz, M Planck, A Sommerfeld, M Laue, M Born, J Larmor, A Eddington, P Debye, P Langevin,
T Levi-Civita --- support the theory, and most of them have made valuable contributions to it.
[\ldots]

I have been accused of running a tasteless advertising campaign for the theory
of relativity. But I can say that all my life I have been a friend of well-chosen, sober
words and of concise presentation.''

\section{1921. ``The Meaning of Relativity''}

In 1921, Einstein was invited to Princeton and delivered there a
course of lectures that make up the book ``The Meaning of
Relativity'' [82]. In this book, he described for the first time,
with maximum exposure to the public and unambiguously, what he
understood by the equivalence of energy and mass. His equations
(41) -- (43) give expressions for the components of the
energy-momentum 4-vector of a body in terms of its mass and
velocity. Equation (44) gives an expression for the energy of a
body in terms of its mass: $E_0=mc^2$. In equation (45), he gave
an expression for energy at a low velocity $q$: $E=m + mq^2/2 +
3mq^4/8 +$\ldots  (in units in which $c=1$.) The text between
equations (44) and (45) reads: ``Mass and energy are therefore
essentially alike; they are only different expressions of the same
thing. The mass of a body is not constant; it varies with changes
in its energy.'' Then follows a footnote about energy release in
radioactive decays: ``The equivalence of mass at rest and energy
at rest which is expressed in equation (44) has been confirmed in
many cases during recent years. In radio-active decomposition the
sum of the resulting masses is always less than the mass of the
decomposing atom. The difference appears in the form of kinetic
energy of the generated particles as well as in the form of
released radiational energy.''

Three aspects deserve our attention in these statements. First, while giving a
clear definition of mass in the equations as a velocity-independent quantity,
the term ``mass at rest'' is used for it, which implies that mass depends on
velocity. Second, there is no explicit statement that mass changes only when
the energy of a body changes, but not its velocity. Third, the ambiguous
statement that mass and energy are ``only different expressions of the same
thing,'' even though mass is a relativistic invariant, i.e., a four-dimensional
scalar, while energy is the fourth component of a four-dimensional vector. It
is possible that these rather imprecise words accompanying perfectly
precise formulas are the reason why many readers still fail to see in
[82] a clear-cut statement in favor of $E_0=mc^2$ and against
$E=mc^2$.

A small popular-science brochure deserves being mentioned here: ``Relativity
theory'' [83], whose author, I Leman, expressed his gratitude to Einstein
for valuable advice. He spoke of his awe
for the profundity and elegance of
Minkowski's ideas and emphasized the enormous amounts of energy stored in matter
as its mass.

\section{1927--1935}

\subsection*{1927}

In 1927, several conferences were dedicated to the bicentennial of the death of Isaac
Newton. Einstein
marked the occasion with a number of publications. He wrote in [84]:

``Newton's teaching provided no explanation for the highly remarkable fact that both the
weight and the inertia of a body are determined by the same quantity (its mass).
The remarkableness of this fact struck Newton himself.''

By 1927, mostly through the work of Einstein, it became clear that
the inertia and the weight of a moving particle are determined not
by its mass but by its energy $E$ and the quantity $p_\mu
\,p_\nu/E$, where $p_\mu$ is energy-momentum vector. In the
Newtonian limit, both are reduced to the rest energy, i.e., to
mass. Such is the simple explanation provided by relativity theory
of the equality of the inertial and gravitating masses in
Newtonian mechanics.

However, we see that Einstein continued to use the old nonrelativistic terminology.

\subsection*{1928}

In the paper ``Fundamental concepts of physics and their most
recent changes'' [85], Einstein formulated his attitude to the
problem of causality in quantum mechanics:

``Thus the field theory shook the fundamental concepts of time, space and matter.
But upon one column of the edifice it made no assault: on the hypothesis of causality.
From some single condition of the world at a given time, all other previous and subsequent
conditions uniquely follow based upon the laws of of nature.

Today, however, serious doubts have emerged about the law of causality thus understood.
This is not to be charged to the craving for new sensations on the part of the learned,
but to the momentum of facts which seem irreconcilable with a theory of strict causality.
It seems at this time as if the field, considered as a final reality, does not make proper
allowance for the facts of radiation and atomic structure. We reach here a complication
of questions with which the modern generation of physicists
is struggling in a gigantic display of intellectual power.''

This problem was solved twenty years later in Feynman's two papers on quantum
electrodynamics
(see below),
but Einstein failed to notice it. This may have been caused by Einstein's
belief that all of quantum physics violated causality.

\subsection*{1929}

In his article for the Encyclopedia Britannica [86], Einstein described the
four-dimensional spacetime continuum but wrote not a word about Minkowski and the
energy--momentum four-dimensional space.

In his speech at the ceremony in honor of the 50th anniversary of Planck's presentation of
his doctoral dissertation --- at which Einstein received the Planck Medal --- he returned to
the problem of causality in quantum mechanics.
He wrote that even though he was deeply convinced that theory would not stop at
the subcausality level and would ultimately reach the supercausality in the
sense discussed by him earlier, he was impressed by the work of the younger
generation of physicists on quantum mechanics,
and that he regarded this theory as a correct one. He only mentioned
that restrictions resulting in the {\em statistical} nature of its
laws should be eliminated with time [87].

\subsection*{1934 -- 1935}

On December 29, the Pittsburgh Post-Gazette published an interview with Einstein
under the heading ``Atom energy hope is spiked by Einstein'' [88].

In December 1934, Einstein read to the joint session of the American Mathematical Society,
the American Physical Society, and the American Society for the Advancement of Science a lecture
entitled ``Elementary derivation of the equivalence of mass and energy.'' This lecture was
published in 1935 in the Bulletin of the American Mathematical Society [89].

The
challenge Einstein
set himself was to prove that mass and energy
are equivalent,
on the basis of only three assumptions:

``In the following considerations,
except for the Lorentz transformation, we will depend only on the assumption of
the conservation principles for impulse and energy.''

In its first pages, Einstein introduces the velocity 4-vector, and by multiplying it by
mass $m$, obtains the 4-vector whose spatial components --- in his opinion --- can naturally
be regarded as momentum and the time component as energy:

``Here it is natural to give it directly the meaning of energy,
hence to ascribe to the mass-point in a state of rest the {\it
rest-energy $m$} (with the usual time unit, $mc^2$).

Of course, \ldots in no way is it shown that this impulse satisfies the
impulse-principle and this energy the energy-principle\ldots

Furthermore, it is not perfectly clear as to what is meant in
speaking of the {\it rest-energy}, as the energy is defined only
to within an undetermined additive constant\ldots

What we will now show is the following. If the principles of conservation of
impulse and energy are to hold for all coordinate systems which are connected
with one another by the Lorentz transformations, then impulse and energy are
really given by the above expressions and the presumed equivalence of mass and
rest-energy also exists.''

And he undertook to prove that conservation laws indeed hold
for the 4-momentum that he considered. To achieve this, he calculated the energies and
momenta of two particles before and after their collision in different Lorentz reference
frames and concluded:

``The rest-energy changes, therefore, in an inelastic collision (additively) like the
mass. As the former, from the nature of the concept, is determined only to within
an additive constant, one can stipulate that $E_0$ should vanish together with $m$.
Then we have simply
$E_0 = m$,
which states the principle of equivalence of inertial mass and rest-energy.''

It is worthy of note here that in this lecture, Einstein never mentioned
Noether's
theorem [75],
which implies that the conservation of the 4-momentum
and the Lorentz invariance follow from symmetry properties
of the Minkowski space-time. He preferred to derive the properties
of the 4-momentum by considering two-body collisions in the three-dimensional space
and to independently assume the Lorentz invariance and conservation of energy and momentum.

On May 4, 1935, he published an obituary in {\it The New York
Times} entitled ``The late Emmy Noether'' [90], where he spoke of
his high opinion of her contributions to mathematics but failed to
mention her theorem that is of such importance in physics. A
self-consistent presentation of conservation laws on the basis of
the symmetries of space -- time in the spirit of Noether was given
for the first time in 1941 by L D Landau and E M Lifshitz in their
``Field theory'' (see below.)

In the same year, 1935, another famous paper was published [91],
written in collaboration with N Rosen and B Podolsky on the interpretation of
measurements in quantum mechanics.

\section{1938 -- 1948. Atomic bomb}

\subsection*{1938}

In 1938, the famous science-popularizing book was published, ``The Evolution of
Physics'' [92], written by Einstein and his young assistant Leopold
Infeld. The authors often returned to the concept of mass on its pages. The section
``One clew remains'' in chapter I
``The rise of the mechanical view''
introduced the concepts of inertial and gravitating masses and
described their equality as
a thread
leading the way to general relativity. In the section ``Relativity and mechanics'' of
chapter III ``Field, relativity,'' the authors introduced the concept of rest mass:
``A body at rest has a definite mass, called rest mass.'' Then they
wrote: ``radiation traveling through space and emitted from the sun contains energy and
therefore has mass;'' and a bit later:
``According to the theory of relativity, there is no essential distinction between
mass and energy. Energy has mass and mass represents energy. Instead of two
conservation laws we have only one, that of mass-energy. This new view proved very
successful and fruitful in the further development of physics.''
One might justly think that this statement is an adequate `verbal' equivalent
of the formula $E=mc^2$ and is incompatible with the formula $E_0=mc^2$.

In the section ``General relativity and its verification'' of the
same chapter III, we read that the elliptical orbit of Mercury precesses,
completing
a full cycle around the Sun in three million years. This
precession of Mercury's perihelion is caused by relativistic properties of the
gravitational field. The next section says this:

``We have two realities: matter and field. [\ldots]
But the division into matter and field is, after the recognition of the
equivalence of mass and energy, something artificial and not clearly defined. Could we not
reject the concept of matter and build a pure field physics?''

The creation of relativistically invariant quantum electrodynamics at the junction of the 1940s and 1950s, and
later of the quantum field theory of the electroweak and strong interactions, as
well as various models of the so-called grand unification of all interactions,
can be regarded as the implementation of Einstein's dream of a unified field
theory. However, all these theories are based not only on the theory of relativity but also
on quantum mechanics, whose probabilistic interpretation was unacceptable to Einstein,
who insisted that ``God
does not play dice.'' It was owing precisely to quantum mechanics that
matter was not expelled from these theories but rather became their foundation.
This is seen especially clearly in the language of Feynman diagrams, in which real
particles (including photons) represent matter and virtual particles represent
force fields (see below).

The concluding chapter IV entitled ``Quanta'' is a story about quantum mechanics. The section
``The quanta of light'' tells the reader that light consists of grains of energy --- light
quanta, or photons. The section
``The waves of matter'' emphasizes the similarity between
photons and electrons in the combination of wave and corpuscular properties.
``One of the most fundamental questions raised by recent advances in science is how to
reconcile the two contradictory views of matter and wave.''
The authors are just a stone's throw from conceding that the photon is just as much a
particle of matter as the electron is. However, at the end of book, they say:

``Matter has a granular structure; it is composed of elementary particles,
the elementary quanta of matter. Thus, the electric charge has a granular structure
and --- most important from the point of view of the quantum theory --- so has energy. Photons
are the energy quanta of which light is composed.''

Light is therefore identified with energy and becomes an antithesis of matter. Could it be
that this identification and this opposition constitute one of the roots of the
mass -- energy confusion?

\subsection*{1939}

On August 2, 1939, Leo Szilard persuaded Einstein to write the famous letter to
President F D Roosevelt warning that ``\ldots the element uranium may be turned
into a new and important source of energy\ldots'' [88].

\subsection*{1941. Landau and Lifshitz}

The first Russian edition of Landau and Lifshitz's ``The theory of
fields'' [93] appeared in 1941. In \S 10 ``Energy and momentum''
(it became \S 9 in subsequent editions of the volume), they
introduced the energy-momentum 4-vector and its square equal to
mass squared, and discussed rest energy, although did not denote
it by $E_0$. The nonadditivity of mass in relativity theory was
mentioned as the nonconservation of mass. All conservation laws in
this book were consistently obtained from the symmetry properties
of space--time in accordance with Noether's theorem. However, it
is very unlikely that Einstein read Russian textbooks. He likewise
missed the publication of the translation into English in 1951
[94].

\subsection*{1942}

In 1942, P G Bergmann's book was published [95] with a foreword by Einstein, which
said, among other things, that:

``This book gives an exhaustive treatment of the main features of the theory of relativity
which is not only systematic and logically complete, but also presents adequately its
empirical basis.

\ldots Much effort has gone into making this book logically and pedagogically satisfactory,
and Dr. Bergmann has spent many hours with me which were devoted to this end.''

In chapter VI, we read:

``\ldots relativistic kinetic energy equals
\begin{equation*}
E= \frac{mc^2}{\sqrt{1-u^2/c^2}} + E_0, \tag{6.17}
\end{equation*}
where $E_0$ is the constant of integration\dots
\begin{equation*}
T= mc^2\left[\left(1-\frac{u^2}{c^2}\right)^{-1/2} -1\right]. \tag{6.20}
\end{equation*}

\ldots The quantity $mc^2$ is called the `rest energy' of a particle, while $T$ is its
`relativistic kinetic energy'. ''

It is not clear to me why the integration constant $C$ had to be denoted by $E_0$.
Neither do I understand why the `relativistic kinetic energy' was denoted by two
different symbols $E$ and $T$. Could it be that a misprint crept in and Eqn (6.17) is the
total, not kinetic, energy? Immediately following it is this text:

``Relation between energy and mass. The ratio between the momentum and the mass, the
quantity $\mu$ , is often called `the relativistic mass' of a particle, and $m$ is
referred to as `the rest mass.' The relativistic mass is equal to the total energy
divided by $c^2$, and likewise the rest mass is $(1/c^2)$ times the rest energy. There
exists, thus, a very close correlation between mass and energy which has no parallel in
classical physics.''

We thus see that the additive constant in the expression for energy and the dependence of
mass on velocity survived in this book. Also retained was the ambiguity connected with
the
definition of the relativistic kinetic energy, which dates back to a 1917 paper
[61].

It looks as if all of this, including the use of the
term ``relativistic mass,'' reflected Einstein's views.

\subsection*{1945}

On August 6, 1945, an atomic bomb was dropped on Hiroshima; another was dropped on
Nagasaki on August 9.

In September, the British magazine `Discovery' published photographs of the
first atomic test explosion on July 16, 1945 and two papers, ``The Progress of
Science --- We enter the New Age'' and ``The Science behind the Atomic Bomb.''
The latter mentioned, in the chronology of the atomic physics discoveries,
``1905. Einstein's special relativity theory demonstrated the equivalence of
mass and energy.'' However, Einstein's photograph was not among the 25
accompanying portraits of scientists from Becquerel to Oppenheimer [96].

In September 1945, the book ``Atomic energy for military purposes'' by H D
Smyth was published [97]. The Introduction said, in the section
``Conservation of mass and of energy'':

``1.2 There are two principles that have been cornerstones of the structure
of modern science. The first --- that matter can be neither created nor destroyed but
only altered in form --- was enunciated in the eighteenth century and is familiar to
every student of chemistry; it has led to the principle known as the law of
conservation of mass. The second --- that energy can be neither created nor destroyed
but only altered in form emerged in the nineteenth century\ldots;
it is known as the law of conservation of energy.

1.3 \ldots but it is now known that they are, in fact, two phases of a single principle
for we have discovered that energy may sometimes be converted into matter and matter into
energy.''

The section ``Equivalence of mass and energy'' said this:

``1.4 One conclusion that appeared rather early in the development of the theory
of relativity was that the inertial mass of a moving body increased as its speed
increased. This implied an equivalence between an increase in energy of motion of a
body, that is, its kinetic energy, and an increase in its mass. \ldots He [Einstein]
concluded that the amount of energy, $E$, equivalent to a mass, $m$, was given by the
equation
$$ E = mc^2 \;,$$
where $c$ is the velocity of light. If this is stated in actual numbers, its startling
character is apparent.''

In these passages, the following deserves our attention:

1. Matter is identified with mass.

2. The law of conservation of momentum is not mentioned, although mass conservation cannot
be understood without it.

3. It is stated that mass increases with velocity.

4. The rest energy and the formula $E_0=mc^2$ are not mentioned.

We also note that H D Smyth was the chairman of the physics department of Princeton University.

\subsection*{1946}

On July 1, 1946, {\it Time} magazine had Einstein's portrait on
its front cover against the background of a nuclear mushroom cloud
with $E=mc^2$ written on it [88].

In 1946, Einstein published two papers on the equivalence of mass and energy:
``Elementary derivation of the equivalence of mass and energy'' [98]
and ``$E=mc^2$: The most urgent problem of our time'' [99].

In the first of them, he partly changed the proof given in 1905 [17]:
the body at rest does not emit radiation but absorbs it; he uses the formulas of
conservation not of energy but of momentum; formulas for the transformation of
energy and momentum of radiation are not used, but instead Einstein uses the
known angle of aberration of stellar light caused by the motion of the Earth:
$\alpha = v/c$. As a result, Einstein obtains the increment to the mass of the
body $M^{\prime}- M = E/c^2$, where $E$ is the energy of the absorbed radiation, and
concludes:
``This equation expresses the law of the equivalence of energy and mass.
The energy increase $E$ is connected with the mass increase $E/c^2$.
Since energy according to the usual definition leaves an additive constant
free, we may so choose the latter that $E=Mc^2$.''

It is obvious from the derivation that $E$ here stands for the
rest energy of the body.

Einstein does not explain why the rest energy of the body is defined up to a constant.

In his brief popular-style article [99], Einstein first described the law of
the conservation of energy using the kinetic and
potential energy of a pendulum as an example,
and then proceeded to deal with the conservation of mass:

``Now for the principle of the conservation of mass. Mass is defined by the
resistance that a body opposes to its acceleration (inert mass). It is also
measured by the weight of the body (heavy mass). That these two radically different
definitions lead to the same value for the mass of a body is, in itself, an
astonishing fact. According to the principle --- namely, that masses remain unchanged
under any physical or chemical changes --- the mass appeared to be the essential
(because unvarying) quality of matter. Heating, melting, vaporization, or combining
into chemical compounds would not change the total mass.

Physicists accepted this principle up to a few decades ago. But it proved
inadequate in the face of the special theory of relativity. It was therefore merged
with the energy principle --- just as, about 60 years before, the principle of the
conservation of mechanical energy had been combined with the principle of the
conservation of heat. We might say that the principle of the conservation of
energy, having previously swallowed up that of the conservation of heat, now
proceeded to swallow that of the conservation of mass --- and holds the field alone.

It is customary to express the equivalence of mass and energy (though somewhat
inexactly) by the formula $E=mc^2$\ldots''

What deserves our attention in this passage is not only what Einstein clarified
but also what he chose not to explain: namely, that the measure of inertia in
relativity theory is not mass but energy and that the quantity $p_\mu p_\nu/E$
creates and feels the gravitational field (and therefore there is nothing
surprising
about the equality
between
the inertial mass and the gravitational mass in Newtonian
mechanics: both are equal to $E_0/c^2$), that the principle of energy
conservation
holds the field not alone but together with
the conservation of momentum, that the energy and momentum determine the mass and
its conservation and/or nonconservation
jointly, and that the mass is
equivalent to
the rest energy.

In 1949 Einstein published ``Autobiographical Notes'' [100], which open
with these words: ``Here I sit in order to write, at the age of 67, something
like my own obituary.'' So in fact he was writing them in 1946 -- 1947. In these
notes, Einstein made an attempt to tell us what and how he had been thinking about for
many years: ``For me it is not dubious that our thinking goes on
for the most part without use of signs (words) and beyond that to a considerable
degree unconsciously.''

On the creation of general relativity he wrote:

``The possibility of the realization of this program was,
however, dubious from the very first, because the theory
had to combine the following things:

(1) From the general considerations of special relativity
theory it was clear that the inert  mass
of a physical system increases with the total energy (therefore,
e.g., with the kinetic energy).

(2) From very accurate experiments (specially from the
torsion balance experiments of E\"{o}tv\"{o}s) it was
empirically known with very high accuracy that the
gravitational mass of a body is exactly equal to its inert mass.''

These words can be interpreted, if one so wishes, as a statement
that the formula $E=mc^2$ not only follows from special (partial)
relativity theory but is also the cornerstone of general
relativity.

\subsection*{1948}

In June 1948, Einstein wrote about the thorny question of mass
for the last time. In a letter to
L Barnett,
author of the book ``The Universe and Dr. Einstein'', he wrote  [101]:

``It is not good to introduce the concept of mass $M=m/\sqrt{1-v^2/c^2}$ of a moving
body for which no clear definition can be given. It is better to introduce no other mass
concept than the ``rest mass'' $m$. Instead of introducing $M$ it is better to mention the
expression for the momentum and energy of a body in motion.''

\section{1949. Feynman diagrams}

In 1949, Feynman published ``The theory of positrons'' [102] and
``Space-time approach to quantum electrodynamics'' [103]. These
papers put quantum electrodynamics into a form that was completely
compatible with the symmetry of the Minkowski world. In these
papers, he formulated and developed a method known as Feynman
diagrams.

The external lines of the diagrams correspond to real on-shell
particles: for them, $p^2=m^2$, where $p$ is the 4-momentum of a
particle and m is its mass. The internal lines correspond to
virtual particles that are off-shell: for these, $p^2\neq m^2$.
Antiparticles look like particles that move backwards in time. All
particles --- both massive and massless --- are described in the
same manner, with a single difference: $m=0$ is assumed for the
latter. (Virtual photons with positive $p^2$ are called timelike,
and those with negative $p^2$, spacelike.) It goes without saying
that the Feynman diagram method is based on the concept of
invariant mass $m$ that is independent of the velocity of the
particle.

Feynman diagrams drastically simplified
calculations for processes involving elementary particles. They
unified all types of matter, both for real particles and for virtual ones that
replaced fields.

F Dyson, who at the time worked with Feynman, recently recalled
[104]:

``During the time that the young physicists at the Institute for Advanced Study in
Princeton were deeply engaged in developing the new electrodynamics, Einstein was working
in the same building and walking every day past our windows on his way to and from the
Institute. He never came to our seminars and never asked us about our work. To the end of
his life, he remained faithful to his unified field theory.''

We know that his famous aphorism---``God is subtle but He is not
malicious''---was engraved above the fireplace at the Institute
for Advanced Study where Einstein worked. One cannot help
recalling his other pronouncement: ``I have second thoughts. Maybe
God {\it is} malicious'' [105].

\section{1952 -- 1955. Last years}

\subsection*{1952}

In 1952, Einstein published a new edition of his popular-science
book ``Relativity, The Special and the General Theory, A Popular
Exposition'' [106], first published in 1917 [61]. For this new
edition, he wrote a special appendix, entitled ``Relativity and
the problem of space,'' in order ``to show that space-time is not
necessarily something to which one can ascribe a separate
existence, independently of the actual objects of physical
reality\ldots In this way the concept of `empty space' loses its
meaning.'' With these words, Einstein was referring not only to
general relativity but also to special relativity theory. The
concept of virtual particles was perhaps alien to him.

\subsection*{1954}

Einstein's foreword to Jammer's book ``Concepts of space'' [107] may contain a
clue to what prevented Einstein from regarding the photon as a material object:

``Now as to the concept of space, it seems that this was preceded by the psychologically
simpler concept of place. Place is first of all a (small) portion of the earth's surface
identified by a name. The thing whose `place' is being specified is a `material
object' or body.''

From this standpoint, any particle, no matter how light,
is a material object while a
strictly massless particle is not.

\subsection*{1955}

In 1895, the 16-year-old Einstein wrote his first scientific essay [108] on the
propagation of light through the ether.

In 1955, in his last autobiographic notes [109], he recalled that at that time a
thought experiment started to puzzle him:

``If one were to pursue a light wave with the velocity of light, one would be confronted
with a time independent wave field. Such a thing doesn't seem to exist, however! This was
the first childlike thought-experiment concerned with the special theory of relativity.''

Thought experiments played an important role in Einstein's research during all his life.

Einstein died on April 18, 1955. A month before his death, Leopold Infeld gave a
talk in Berlin at a meeting that celebrated the 50th anniversary of relativity
theory [110]. He named the dependence of mass on velocity as the first of the three
experimental confirmations of special relativity theory. The baton of ``relativistic
mass'' was passed on to future generations.

\section{Born, Landau, Feynman}

\subsection*{Born's books}

An important role in this passing of the baton belongs to Max Born's book
``Einstein's theory of relativity.'' An outstanding physicist, one of the creators of
quantum mechanics, Born did very much to help spread relativity theory. The first edition
of his book appeared in 1920 [111] (its Russian edition was published in 1938
[112]). The next edition [113] appeared after Einstein's death in 1962
(and its translation into Russian [114] in 1964 and 1972).
Unfortunately, both these editions, which greatly influenced how physics was taught in the
20th century, state without any qualifications that the increase in the mass of a body
when its velocity increases is an experimental fact.
It is also asserted in [115, 116].

In 1969 --- a year before passing away --- Born published his
correspondence with Einstein [117], which lasted from 1916 till
1955. Not even one among more than a hundred letters touches on
the aspect of the [in]dependence of mass on velocity. The
correspondence was translated into English; its latest edition was
published in 2005 [118] with a detailed foreword, which also
ignored the mass controversy.

\subsection*{Landau and Rumer brochure}

I mentioned above that the Landau -- Lifshitz book ``Field theory'' [93]
was the first monograph on relativity theory in world literature that
consistently applied the idea that the mass of a body is independent of its
velocity. It is all the more incomprehensible why in their popular brochure
``What is relativity theory?'' [119, 120], Landau and Rumer
chose for the first introduction into the theory the statement that mass is a
function of velocity and that this is an experimental fact. In the third edition
of this brochure published in 1975, Yu B Rumer added ``Pages of reminiscences
about L D Landau,'' where he quoted a jocular characteristic of the brochure
given by Landau himself: ``Two con men trying to persuade the third one that for
the price of a dime he would understand what relativity theory is.''

\subsection*{The Feynman Lectures}

The magnificent lectures on physics that Feynman
gave to students of Caltech in 1961 -- 1964
[121]
instilled
a love
for physics in the hearts of millions of readers around the world
(see, e.g. [122]). They teach readers to think independently and
honestly. Alas, these lectures never
mention the Feynman diagrams that he invented in 1949 [102, 103] and which
brought him the Nobel prize in 1965. Furthermore, the entire relativity theory is
introduced in these lectures through the formula $E=mc^2$, not through the concept of the
Lorentz-invariant mass on which Feynman diagrams are based.

Feynman states already in the first chapter that the dependence of
the mass of a body on its velocity is an experimental fact, in the
fourth he says that Einstein discovered the formula $E=mc^2$, and
in the seventh that mass is the measure of inertia. In chapter 15,
we meet the formula $m=m_0/\sqrt{1-v^2/c^2}$ and Feynman discusses
the consequences of the ``relativistic increase of mass''; in
chapter 16, he derives this formula. This chapter ends with the
words:

``That the mass in motion at speed $v$ is the mass $m_0$ at rest
divided by $\sqrt {1-v^2/c^2}$, surprisingly enough, is rarely
used. Instead, the following relations are easily proved, and turn
out to be very useful: $E^{\,2}-P^2 c^2 = {M_0}^2c^4$ and $Pc = E
v/c$.'' (The original notation used by Feynman is retained in this
quotation.)

Even in Chapter 17, where Feynman introduces four-dimensional space--time and uses units
in which $c=1$, he continues to speak of the rest mass $m_0$, not simply of the mass $m$.

In the course of 2007, I e-mailed a question to a number of Feynman's former
students, assistants and co-authors. Not one of them was able to recall even a
single occasion when Feynman used the notion of relativistic mass or the
formula $E=mc^2$ in discussions he had with them. Nevertheless, several millions
of readers of his lectures firmly believe that mass is a function of velocity.
Why would the great physicist who gave us the language of Feynman diagrams place
the notion of velocity-dependent mass at the foundation of his Feynman lectures?

Perhaps we can find an answer to this question in Feynman's Nobel lecture
[123]. He described there numerous `blind alleys'
in which he
had been
trapped while on his way to constructing quantum electrodynamics, but still
expressed the firm belief that ``many different physical ideas can describe the same
physical reality.''

Thus he wrote about the idea of an electron moving backwards in time:
``it was very convenient, but not strictly necessary for the theory because
it is exactly equivalent to the negative energy sea point of view.''
However, without a time-reversed electron, there would be no Feynman diagrams, which
introduced order and harmony into huge areas of physics.

\section{Epilogue}

Why is it that the weed of velocity-dependent mass is so resistant?
First and foremost, because it
does not lead to immediate mistakes as far as arithmetic
or algebra are concerned. One can introduce
additional `quasi-physical variables' into any self-consistent theory by
multiplying true physical quantities by arbitrary powers of the speed of light.
The most striking example of such a `quasi-quantity' is the so-called
`relativistic mass.' If calculations are done
carefully enough, their results
should be the same as in the original theory. In a higher sense, however, after
the introduction of such `quasi-quantities,' the theory is mutilated because its
symmetry properties are violated. (For example, the relativistic mass is only one
component of
a 4-vector, while the other three components are not even mentioned.)

Some other explanations of the longevity of relativistic mass can
be given here. The formula $E=mc^2$ is `simpler' than the formula
$E_0=mc^2$ because the additional zero subscript that requires
explanation is dropped. The energy divided by $c^2$ indeed has the
dimensionality of mass. Intuition based on conventional everyday
experience slips in a hint that the measure of inertia of a body
is its mass, not its energy, and this prods one to `drag' the
nonrelativistic formula ${\bf p} = {m}{\bf v}$ into relativity
theory. The same intuition suggests, with hardly less insistence,
that the source of gravitation is `our' mass, not an `alien'
quantity $p_\mu\,p_\nu/E$. Everyday experience rebels particularly
strongly against the idea of treating light as a type of matter.
The arguments given above may explain the `Newtonian bias' of an
ordinary person, let us say `a pedestrian.' However, it would be
too flippant to attribute them to such a great physicist as
Einstein. Indeed, it was Einstein who introduced the concept of
rest energy $E_0$ into physics and wrote about $E_0=mc^2$ far more
often than about $E=mc^2$. Still, one thing remains unexplained:
why was it that during the half-century of discussing the relation
between mass and energy, Einstein never once referred in either
his research publications or his letters to the formula $E^2-{\bf
p}^2 c^2=m^2 c^4$, which defines the Lorentz-invariant mass?

It is possible that the formulation of the total equivalence of energy and mass
reflected Einstein's absolute reliance on his powerful intuition. It was without
a doubt his confidence in his own intuition that resulted in his rejection of quantum
mechanics. One feels that he perceived the concept of electromagnetic potential not
only
with his mind but
with his entire body. And he `felt' the wave function to be very much like
the electromagnetic wave.
His resistance to quantum mechanics
prevented Einstein's world line
from meeting Feynman's world line
in the space of ideas --- in the noosphere, so to speak. As a consequence,
Einstein refused to accept the photon as a
particle of matter and continued to treat
it as a quantum of energy.

\section{Conclusion}

When shown an art exhibition in Moscow Manege in 1962, Nikita Khrushchev
(1894 -- 1971) rudely attacked the sculptures of Ernst Neizvestny. When Khrushchev
died, his children requested Neizvestny to create a memorial sculpture at the
grave of their father. The main part of this memorial
consists of two vertical
marble slabs, one white, the other black, whose protrusions penetrate each other. These slabs in a
way symbolize the good and the evil.

The history of the confrontation of two concepts of mass in the 20th century resembles
this
sculpture. Here the light and the darkness were fighting each other in the minds of the
creators
of modern physics.

In the world of opinions, pluralism is considered to be
politically correct. To insist on a single point of view is
thought to be a manifestation of dogmatism. A good example of
fruitful pluralism is the wave-particle duality in quantum
mechanics. But there are cases in which a situation is ripe for
establishing unambiguous terminology. The relation between energy
and mass is more than ripe for this. It is high time we stopped
deceiving new generations of college and high school students by
inculcating into them the conviction that mass increasing with
increasing velocity is an experimental fact.

\section*{Postscriptum.
In memory of J A Wheeler}

This review was already completed when I received the sad news that John Archibald
Wheeler, an outstanding physicist and teacher who accomplished so much for
establishing the spacetime interpretation of relativity theory and of the concept of
Lorentz-invariant mass, died on 13 April 2008, at the age of 96. I dedicate
this paper to his memory.

\section*{Acknowledgments} I am grateful to B L Okun, M B Voloshin,
and V I Kisin for their invaluable help in preparing this paper. I
also appreciate helpful discussions with A A Abrikosov, T B
Aksent'eva, A A Alekhina, B L Altshuller, J M Bardeen, T Basaglia,
S M Berman, S I Blinnikov, B M Bolotovsky, L M Brown, D K
Buchwald, T L Curtright, Yu Danoyan, A D Dolgov, M A Gottlieb, M D
Godfrey, Ya I Granovsky, E G Gulyaeva, D R Hofstadter, J D
Jackson, M Janssen, C Jarlskog, O V Kancheli, M Karliner, V I
Kogan, Ya S Kim, C Quigg, G L Landsberg, L Yu Mizrahi, V A
Novikov, L I Ponomarev, P S Prokof'ev, F Ravndal, M Sands, S G
Tikhodeev, K A Tomilin, V P Vizgin, M I Vysotsky, V R Zoller, G
Zweig. The work was supported by the grants NSh-5603.2006.2,
NSh-4568.2008.2, RFBR-07-02-00830-a.

\end{document}